\documentclass{vldb}
\usepackage{epsfig,endnotes,listings,xcolor}
\usepackage{epstopdf}
\usepackage{graphicx}
\usepackage{subfigure}
\usepackage{array}
\usepackage{amssymb}
\usepackage{amsmath}
\usepackage{multirow}
\usepackage{algorithm}
\usepackage{algpseudocode}
\usepackage{url}
\usepackage{hyperref}

\begin{document}

  \title{PFO: A Parallel-Friendly High Performance System for Online Query and Update of Nearest Neighbors}
  \author{
  %
  %
  \numberofauthors{2}
  \alignauthor
  Nan Zhu\qquad Wenbo He \qquad Xue Liu\\
  \affaddr{McGill University}\\
  \affaddr{Montreal, QC, Canada}\\
  \email{\{nan.zhu, wenbohe, xueliu\}@cs.mcgill.ca}
  \alignauthor
  Yu Hua\\
  \affaddr{Huazhong University of Science and Technology}\\
  \affaddr{Wuhan, Hubei, China}\\
  \email{csyhua@hust.edu.cn}
  }

\maketitle

\begin{abstract}
  Nearest Neighbor(s) search is the fundamental computational primitive to tackle massive dataset. Locality Sensitive Hashing (LSH) has been a bracing tool for Nearest Neighbor(s) search in high dimensional spaces. However, traditional LSH systems cannot be applied in online big data systems to handle a large volume of query/update requests, because most of the systems optimize the query efficiency with the assumption of infrequent updates and missing the parallel-friendly design. As a result, the state-of-the-art LSH systems cannot adapt the system response to the user behavior interactively.

  In this paper, we propose a new LSH system called PFO. It handles query/update requests in RAM and scales the system capacity by using flash memory. To achieve high streaming data throughput, PFO adopts a parallel-friendly indexing structure while preserving the distance between data points. Further, it accommodates inbound data in real-time and dispatches update requests intelligently to eliminate the cross-threads synchronization. We carried out extensive evaluations with large synthetic and standard benchmark datasets. Results demonstrate that PFO delivers shorter latency and offers scalable capacity compared with the existing LSH systems. PFO serves with higher throughput than the state-of-the-art LSH indexing structure when dealing with online query/update requests to nearest neighbors. Meanwhile, PFO returns neighbors with much better quality, thus being efficient to handle online big data applications, e.g. streaming recommendation system, interactive machine learning systems.
\end{abstract}

\section{Introduction}
\label{sec:intro}

The recent exponential growth of social media, multimedia, and machine learning applications have produced many instances where high-dimensional feature vectors are used to represent data, and gigantic storage spaces are needed. The searches on a vast amount of vectors to find Nearest Neighbor (NN) is very slow due to the large search range as well as the well-known ``Curse of Dimensionality" in NN search \cite{exactnn1}. Locality Sensitive Hashing (LSH) was proposed in \cite{basiclsh} as a bracing solution for NN search in high-dimensional space. The basic idea of LSH is to reduce the dimensionality of the data points by mapping them into a lower-dimensional space with the specially designed distance-preserving hash functions. Consequently, the hash values of the data points are the same or similar with high probability if they are close to each other in the high-dimensional space. A wide variety of LSH algorithms with many optimization strategies on LSH have been designed to improve the query performance \cite{lsbtree, lshcollisioncounting, sklsh} or memory efficiency \cite{multiprobelsh}.

\begin{table}
\tiny
\centering
\begin{tabular}{| p{1.5cm} | p{1.5cm} | p{1.5cm} | p{1.5 cm} |}
   \hline
    & \textbf{Storage Medium} & \textbf{Parallel Data Access} & \textbf{Online Update} \\ \hline
   Multi-Probe LSH \cite{multiprobelsh} & RAM & Single Thread & No \\ \hline
   LSB-Tree \cite{lsbtree} & Disk & Single Thread & No \\ \hline
   C2LSH \cite{lshcollisioncounting} & Disk & Single Thread & No \\ \hline
   SK-LSH \cite{sklsh} & Disk & Single Thread & No \\ \hline
   PLSH \cite{plsh} & RAM & Distributed & Pause the processing to consume it \\ \hline
   Distributed LSH \cite{distributedlsh} & Disk & MapReduce & No \\ \hline
   Hypercurves \cite{cpugpulsh} & RAM & CPU\&GPU for Construction & No \\\hline
   GPU-LSH \cite{gpuknn} & RAM & GPU for Construction & No \\ \hline
   PFO & Hierarchical Memory System & Multi-Threaded & Parallel-friendly Index and Smart Write Task Dispatching \\ \hline
\end{tabular}
\caption {Analysis of Representative LSH systems and How PFO differentiates itself with them}
\label{tab:lsh_related}
\end {table}

However, we identified several facts that hinder the application of LSH in online big data systems. (1) A challenge we face is the dilemma that we should whether to pursue data access efficiency or storage scalability? For the seek of the efficiency, we prefer to use RAM to implement LSH; but for scalability, usually people adopt disk storage. Achieving both efficiency and scalability is desired for big data applications with online performance requirements, but the state-of-the-art design of LSH systems \cite{plsh, multiprobelsh, sklsh, lshcollisioncounting} cannot achieve both goals. (2) While the existing LSH systems aim to optimize the efficiency of a single query request \cite{lsbtree, lshcollisioncounting, sklsh}, a practical LSH system must be able to handle concurrent queries/updates. It requires that we design a new parallel-friendly indexing structure for LSH systems. Though the prior efforts have been made to place data \cite{plsh} in distributed servers and execute queries by broadcasting requests to all servers or in batch \cite{distributedlsh}, we are in need of new LSH system with indexing structure scaling against the query/update requests which arrive concurrently. (3) The current LSH designs \cite{lsbtree, lshcollisioncounting, sklsh} lack the support of online data update, which is indispensable in many big data systems (e.g., online recommendation systems \cite{tencentrec}). In these systems, user's online browsing or search activities affect the system response. A system needs to remember individual user's input and respond accordingly. Hence, a system needs to update the data dynamically and frequently. The contemporary LSH systems \cite{gpuknn, cpugpulsh} are designed with the assumption of no or infrequent update requests. The other designs \cite{sklsh, plsh} usually pay additional space and/or accuracy cost to maintain the updated version of the index. It is not practical to use LSH in contemporary big data systems if LSH fails to support the online update efficiently. We summarize the existing LSH research work in Table \ref{tab:lsh_related}.

Despite its salient performance when dealing with high dimensional data, the gap between the state-of-the-art LSH research and the requirement of the big data applications prevents these LSH-based systems from being adopted in the more challenging, yet more realistic, online scenario. To bridge the gap, we design \emph{PFO}, A \textbf{P}arallel-\textbf{F}riendly High-Performance System for \textbf{O}nline Query and Update of Nearest Neighbors, utilizing multiple system-oriented optimization strategies to serve massive, concurrent, online requests. Specifically, we made the following contributions:

\begin{itemize}
\item \textbf{Hierarchical Memory} To resolve the dilemma between the query efficiency and system scalability, we adopt a hierarchical memory system in PFO. All query and update requests to PFO are first consumed in RAM memory space to improve the processing speed. In RAM memory space, we employ a data placement strategy to reduce the overhead brought by Garbage Collection. To scale the system capacity, we write data to flash memory and organize data in a read-friendly format to improve the query efficiency.
\item \textbf{Parallel-Friendly Design} We propose an indexing structure called Partitioned Hashing Forest (PHF) to facilitate the parallel data access. By applying the LSH functions to the data points and then to their hash values, we divide the memory space into two levels, called \emph{HashTree} and \emph{Partition} level. With PHF, we partition the data points in PFO with the respect to their location in the high-dimensional space. Each partition can be accessed in parallel without the crossing dependence. Additionally, we have a concurrency management strategy to avoid the synchronization across threads. By eliminating the cross-threads synchronization, we significantly improve the performance when the query and update requests arrive in the system simultaneously.
\item \textbf{Reconstruction-Free Hash Tree} We build a hash tree structure which accepts the online update requests without the necessary to reconstruct itself. With the hash tree, PFO accommodates the online update requests more efficiently than the other data structures which require reshaping the index, e.g. B-Tree \cite{lsbtree}.
\end{itemize}

We implemented our PFO in a prototype and evaluated on 10-cores server with the standard benchmark datasets. We compared the performance of PFO with a state-of-the-art LSH system as well as PFO's variations. Our results show that PFO: 1) offers sub-second query latency with the growing size of the indexed data in the hierarchical memory system; 2) scales processing throughput close to linearly with the increasing number of cores; 3) improves the system throughput up to five times when dealing with the online query/update requests, comparing to other indexing structures. We also validate the accuracy of PFO by comparing with the state-of-the-art LSH-based search algorithm. The results show that PFO converges to the ideal accuracy metric with much fewer hash tables.

The rest of the paper is organized as follows. Section \ref{sec:background} presents the background knowledge, the problem settings and our objectives. Section \ref{sec:storage} goes into the details of PFO's storage system with the hierarchical memory design. In Section \ref{sec:parallel}, we discuss the parallel-friendly indexing structure and the concurrency management module in PFO. Section \ref{sec:online} describes how we use the adaptive hash tree to handle online update requests. Section \ref{sec:database} compares the design philosophy of PFO and the general online database system. Section \ref{sec:evaluation} contains an extensive experimental evaluation. Section \ref{sec:related} points out the defects of the existing LSH-based techniques. Finally, Section \ref{sec:conclusion} concludes the paper with a summary of our findings.

\section{Background and Use Case Study}
\label{sec:background}
In this section, we give a brief introduction to Locality Sensitive Hashing (LSH), Nearest Neighbor (NN) search problem and its variance, Approximate Nearest Neighbor (ANN) search. Then we discuss how real-world applications extends the use of LSH to accelerate the search for ``Nearest Neighbors".

\subsection{Locality Sensitive Hashing}
\label{sec:background:lsh}

Given a dataset $D \subset \mathbb{R}^d$ and a query data point $q$, Nearest Neighbor (NN) search problem aims to find a data point $o$ in $D$ and, allows any other data points $p \in D$, $\lVert \mathbf{p, q} \rVert \geq \lVert \mathbf{o, q} \rVert$, where $\lVert \rVert$ represents the distance between two points. To improve the efficiency of NN search in a high-dimensional space, researchers developed approximate version of NN search algorithms, i.e. Approximate Nearest Neighbor (ANN), to locate $o^* \in D$, where $\lVert \mathbf{o^*, p} \rVert \leq c\lVert \mathbf{o, p} \rVert$, c is the approximate ratio and usually larger than 1.

Locality Sensitive Hashing (LSH) \cite{lsh} is one of the most popular ANN search algorithms, which is defined as follows:

\emph{Definition 1} (Locality Sensitive Hashing) Given a distance $R$, an approximate ratio $c$ as well as two probabilities $p_1$ and $p_2$, a hash function $h: R^d \rightarrow \mathbb{Z}$ is called $(R, c, p_1, p_2)$-sensitive, if

\begin{itemize}
  \item If $\lVert \mathbf{p, q} \rVert \leq R$, then $Pr[h(p1)=h(p2)] \geq p_1$;
  \item If $\lVert \mathbf{p, q} \rVert \geq cR$, then $Pr[h(p1)=h(p2)] \leq p_2$;
\end{itemize}

We guarantee that $c > 1$ and $p_1 \geq p_2$. In practice, we use a \emph{Compound Hash Key} to map the object into a bucket in a LSH hash table. Given a compound hash function $G = (h_1, ...,  h_m)$, the bucket ID  consists of $(h_1(p), ... h_m(p))$. In this paper, we adopt the angular distance between two unit vectors used in \cite{plsh}. The hash function is parameterized by an unit vector $a$, and the hash value of the query object $q$ is $h_a(q) = sign(a . q)$.

Within a single hash table in LSH, the conventional approaches only take the data points with the same hash value as the candidates of the nearest neighbor. This rule is too strict as it filters out too many data points. To mitigate this, we usually use $L$ hash tables in an LSH-based system.  Multiple hash tables in LSH causes significant memory overhead to the LSH-based system. To reduce the number of the required hash tables in LSH, one approach is to take the data points with the ``similar" hash values as the candidates of the nearest neighbor \cite{multiprobelsh, sklsh, lsbtree}. To measure the distance between two compound hash values, we adopt the similar definition as in \cite{lsbtree}:

\emph{Definition 2}. (Distance of Compound Hash Values). Given two compound hash keys $K_1$ and $K_2$, the distance, denoted as $dist(K_1, K_2)$, is defined as follows: $dist(K_1, K_2) = \frac{1}{LLCP(K_1, K_2)}$, where LLCP is the longest length of the common prefix of two compound hash values.

\subsection{Online Nearest Neighbors Search via LSH}
\label{sec:background:scenario}

In this section, we show a real-world scenario for LSH-based systems, where the goal is to get the ``Nearest Neighbors'' of a given query data point, instead of the ``nearest one''. We consider an online recommendation system. Recommendation systems \cite{tencentrec, googlenews} are extensively used in online e-commerce systems and online social networks such as Amazon, Netflix, Twitter, etc. In these systems, operators want to track the click history or users and represent the click history for a user as a vector, i.e. a data point in the vector space. By calculating the similarities between the data points, the recommendation system provides recommendations for users based on the behaviors represented by similar data points. However, paired comparison of similarity in a large dataset (e.g. Google News or Amazon) is costly. As a result, users may not be able to get recommendations in time. To reduce the range of the vectors to be included in similarity calculation, the existing works \cite{googlenews, plsh} propose to use LSH to locate the candidates of the nearest neighbors. Then they take the candidates into the further similarity calculation and output the nearest ones whose similarity with the query vector is beyond a user-defined threshold (i.e. R in Definition 1). Though LSH systems are employed to accelerate the similarity calculation, the legacy systems rely on the periodical and offline computation to construct recommendation model. To suggest recommendations to the users timely, recommendation systems must update the click history of users continuously. In this scenario, the offline systems cannot generate accurate and in-time recommendations to users \cite{tencentrec}. To enable the applications like recommendation systems to work online, we must make the LSH-based systems (1) scale out to handle increasing amount of the data points; (2) provide high throughput for concurrent query/update requests. The design of PFO in this paper focus on these two goals.

\section{Storage of PFO}
\label{sec:storage}

\begin{figure}[t]
  \begin{center}
    \includegraphics[height=3.2cm]{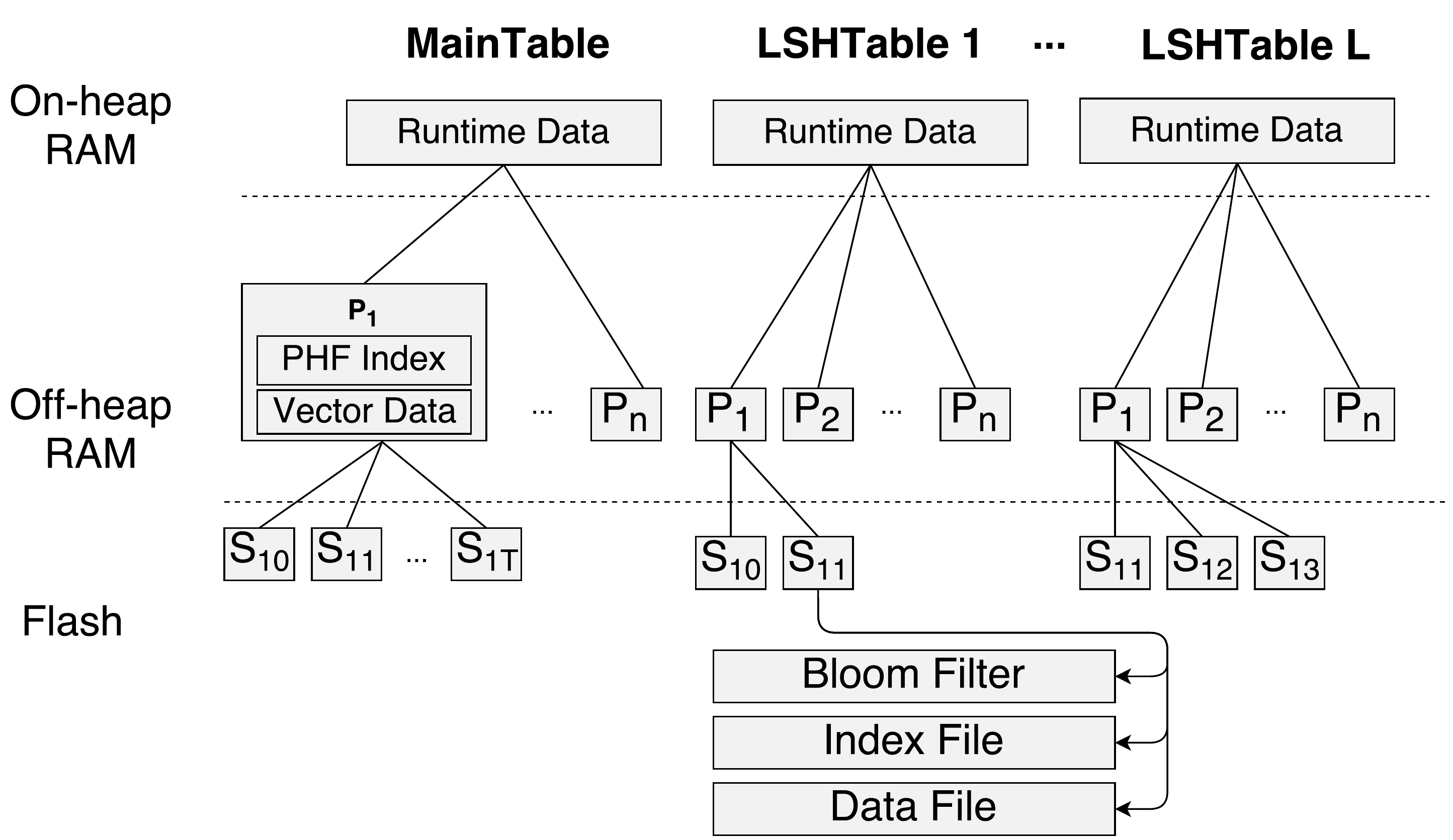}
    \caption{\small{Storage Memory Layer Design: PFO saves data in both RAM and Flash Memory (SSD-based). It consumes all I/O in RAM to improve the performance and saves snapshots ($S_{ij}$) in SSD for increasing the system capacity with the performance guarantee. Additionally, the off-heap space for each table is divided into multiple partitions ($P_i$) for exploit parallelism for processing. }}
    \label{fig:storage}
  \end{center}
\end{figure}

In this section, we present the overview of PFO's storage component, consisting of multiple data tables, each of which locates across hierarchical memory space. This design resolves the dilemma between the efficiency and capacity of LSH systems. Figure \ref{fig:storage} illustrates the storage layers in our design. In the following subsections, we step through the components in the figure. We begin the description with how we organize data points with multiple data tables in Section \ref{sec:storage:tables}. After that, we move forward to the design of hierarchical memory system of PFO in Section \ref{sec:storage:hierarchical_memory}. We list the notations used in this paper in Table \ref{tab:notation}.

\begin{table}[ht]
\tiny
\centering
\begin{tabular}{ | c | p{5cm} | }
   \hline
    \textbf{Notations} & \textbf{Explanation} \\ \hline
   $L$ & the number of hash tables used in LSH schema\\ \hline
   $C$ & the number of LSH hashing functions which are used to locate the data point in partition level  \\ \hline
   $S_{ij}$ & the snapshot of partition $i$ of a hash table locating in flash memory which is saved at moment $j$\\ \hline
   $m$ & the number of bits in a data point's hashing value which is used to locate the hashing tree in $H$ layer \\ \hline
   $t$ & the allowed number of data points residing in the same bucket (except the last level of the hash tree) \\ \hline
   $l$ & the number of slots in the directory node \\ \hline
   $s$ & the size of the data point  in terms of bytes \\ \hline
   $r$ & the record ID of the data point  \\ \hline
   $h$ & hash value of the data point \\ \hline
   $A(q)$ & nearest neighbors candidates of the query data point, the data points in this set whose distance to the query data point is no larger than the user-defined threshold, R, are selected as nearest neighbors \\ \hline
   $o$ & offset of the leaf in off-heap space \\ \hline
   $M$ & maximum length of hash key in PFO in terms of bits \\ \hline
   $h'$ & variation of $h$ by discarding $h$'s' last $i * log_2(l)$ bits, where i is an integer ranging from 0 to the height of the hash tree \\ \hline
   $h'_{max}$ & the maximum length of the existing $h'$ in hash table in terms of bits \\ \hline
   $KL$ & the longest length common prefix of a data point's hash key and the other one \\ \hline
\end{tabular}
\caption {Notations of PFO}
\label{tab:notation}
\end {table}

\subsection{Data Tables in PFO}
\label{sec:storage:tables}

LSH systems usually employ multiple LSH tables to achieve better accuracy for nearest neighbors search. However, the state-of-the-art systems have focused on the optimization of individual LSH tables, ignoring the overall performance and overhead of the systems. For example, in LSH systems described in \cite{lsbtree, sklsh}, the data points are usually represented as high-dimensional vectors. The vectors have multiple copies and are saved in various LSH tables, which incurs additional storage overhead. Even worse, in the LSH systems like \cite{sklsh}, the order of the vectors in an LSH table affects the query response performance. In online update scenarios, the vectors are subject to be frequently updated. Upon updating a vector, the order of the vectors may need to be changed, and additional efforts are required to maintain the index of the vectors in multiple LSH tables, in order to optimize the query response of the LSH systems.

To make an LSH system more efficient and make it suitable to deal with online updates, we adopt a ``MainTable + LSHTables" structure in PFO. As shown in the top line in Figure \ref{fig:storage}, PFO consists of one MainTable and L LSHTables. The MainTable stores the data points (represented as vectors), which are indexed by the unique ID. The LSHTables only store the IDs of the data points. Therefore, PFO (1) saves the storage space by keeping a single copy of the vectors in MainTable while we use multiple LSHTables for better search accuracy; (2) makes update across multiple LSH tables more efficient:

\begin{itemize}
  \item PFO handles query and update requests as follows. When fetching nearest neighbors for the given query data point $q$, we calculate q's hash values with the hash functions associated with L LSHTables and fetch the IDs of the nearest neighbors candidates, say $A(q)$, according to q's hash values from all these tables. We then remove the duplicate IDs and read the corresponding vectors from MainTable with the remaining ones. When adding a new vector to PFO, we first save it in MainTable and then update L LSHTables with its ID according to its hash values in these tables. With this approach, we keep a single copy of all vectors in MainTable and preserve the distance between the vectors with LSHTables.
  \item MainTable is built with the hash function minimizing the hash conflict, e.g. MurmurHash3 \cite{murmurhash3}, providing an O(1) complexity when reading/inserting/deleting elements on average. Since MainTable does not impose any constraint on how the vectors shall be placed in the memory space, we do not have to rebuild the whole hash table for vector updates.
\end{itemize}

\subsection{Hierarchical Memory System}
\label{sec:storage:hierarchical_memory}

As we discussed in Section \ref{sec:background:scenario}, the online environment brings the challenges on both handling capacity and offering fast responsiveness to PFO. Therefore, pure-RAM approaches like E2LSH \cite{basiclsh} and Multi-probe LSH \cite{multiprobelsh}, or pure-disk approaches like SK-LSH \cite{sklsh} are not ideal in the case. In this section, we introduce how we overcome the challenges with Hierarchical Memory System in PFO. The left part of Figure \ref{fig:storage} suggests that each table in PFO locates across three types of memory: \emph{on-heap}, \emph{off-heap} and \emph{flash} memory. Among the three components, on-heap and off-heap are in RAM and flash memory is built with Solid State Disk (SSD).

\subsubsection{RAM}
\label{sec:storage:hierarchical_memory:ram}

To respond quickly to the query/update requests and offer scalable storage capacity simultaneously, we employ both RAM and flash memory in PFO. When handling a query request, we first search the $A(q)$ in RAM before we go to the disk space. When we update a data point, it is added in the RAM. If the RAM is overloaded, the data points will be dumped to flash memory.

Recently, more and more database systems serving a large volume of online user requests are developed in programming languages that rely on automatic Garbage Collection (GC) \cite{jvmgc} to reclaim unused memory space, such as HBase \cite{hbase} and Cassandra \cite{cassandra} serving Facebook and Apple respectively. However, when the program is in significant memory pressure and getting free memory space is tough, GC will bring considerable overhead to the user applications \cite{coordinategc, avoidgc}. In the worse case, GC threads pause the application threads, trying to get free memory space with the best efforts. To minimize the negative effect brought by GC, we divide the RAM space in PFO into on-heap and off-heap spaces, and the latter is not affected in GC.

We save different types of data in on-heap and off-heap spaces. In \textbf{on-heap} part, we only keep the runtime data, which includes system runtime parameters or the handler of the off-heap data in the on-heap space. The runtime data is small and brings ignorable overhead to the program memory. We save the PFO tables in \textbf{off-heap} space. Each table is divided into independent partitions (referred as $P_i$ in Figure \ref{fig:storage}) to facilitate the parallel access to the whole table. Every partition contains the content incurring most of memory cost, \emph{Data} and \emph{Index}. For MainTable, Data is the vectors representing the data points; while for LSHTables, Data are the IDs of the vectors. To locate the IDs (in LSHTable) and vectors  (in MainTable) quickly, we employ a Partitioned Hash Tree (PHF) Index to serve the purpose. As the name suggests, PHF is a set of hash trees, the non-leaf nodes of which are kept in the off-heap memory as Index. (We will leave the detailed introductions to the partitioning algorithm and the PHF index to Section \ref{sec:parallel:indexing}). By saving MainTable and LSHTable in off-heap memory that is not subject to GC, we minimize the chances to trigger GC, therefore, achieve the high system performance of PFO.

\begin{figure}[t]
  \begin{center}
    \includegraphics[height=3.5cm]{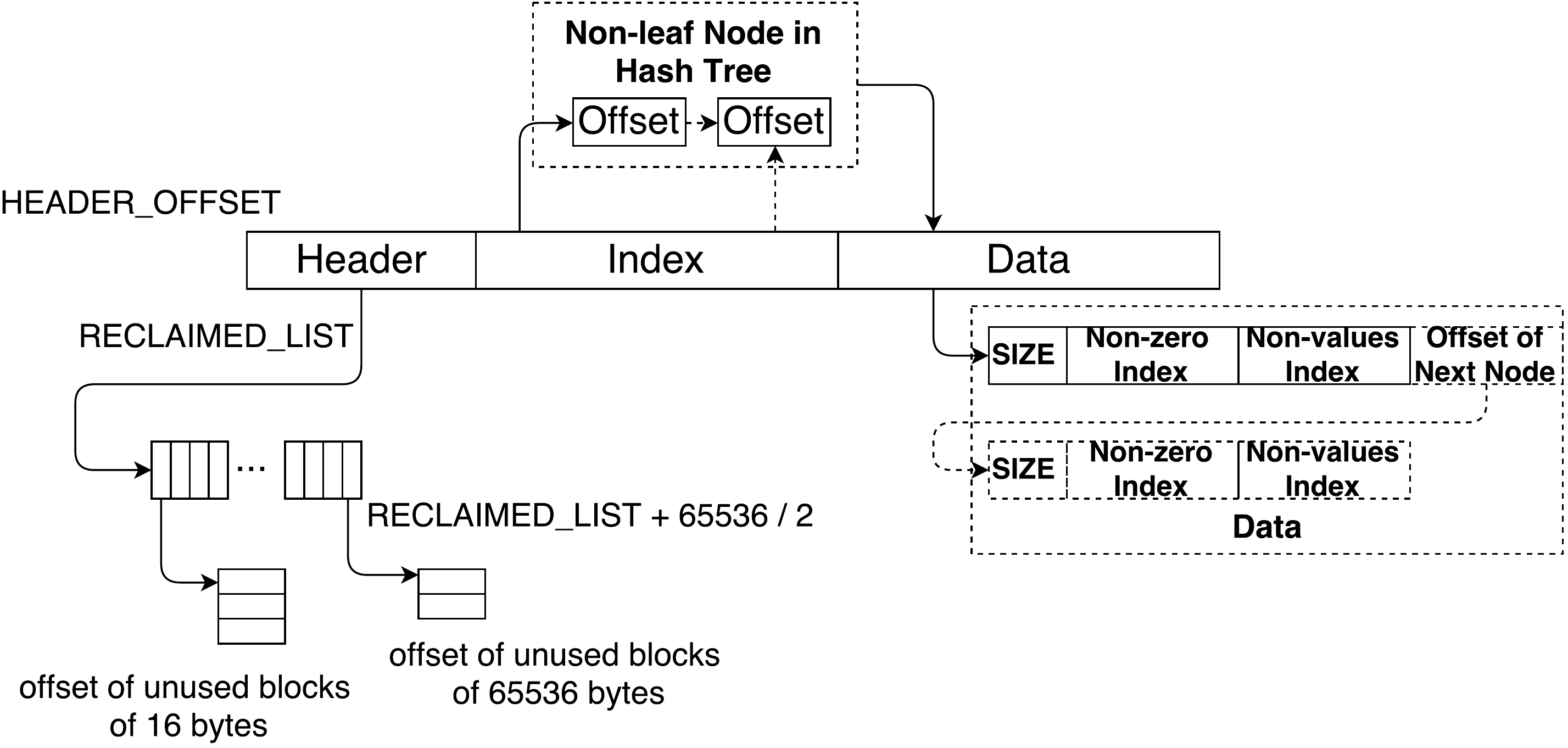}
    \caption{\small{Memory Layout in Off-heap space. In the figure, we only show the layout of MainTable which has vectors in Data segment for simplicity. For LSHTables, the content in Data segment is the IDs of the vectors.}}
    \label{fig:mem_layout}
  \end{center}
\end{figure}

We describe the memory layout in a single partition of the PFO tables in Figure \ref{fig:mem_layout}. By saving the roots of the hash trees in the fixed offset, we start searching $A(q)$ in the hash tree by fetching the root node, step down the levels and eventually locate the leaf node, i.e. the vectors or vector IDs. As a result, we may perform read operations for multiple times for \textbf{\emph{Index}} segment and for each time, we get the offset indicating the next memory block to read. To further reduce the memory cost in MainTable, we save the vector in a compressed format in \textbf{\emph{Data}} segment. We design the format based on the fact that the vectors are mostly sparse in many scenarios, e.g. recommendation system \cite{tencentrec}, machine learning \cite{sparsefeature}. A vector consists of 3 fields, size, non-zero indices and non-zero values. ``size" is the dimensionality of the vector, ``non-zero indices" records the dimensions where the vector has a non-zero value and ``non-zero values" are the non-zero values in the dimensions referred by non-zero indices field. To resolve the hash conflict, we have the offset field in each data point indicating the offset of the next data point with the same hash value. A major challenge for off-heap memory space design comes from the fact that a vector size is subject to change if the update is allowed. As a result, we need to have a fast approach to invalidate the memory space for stale vectors, and reclaim the memory for future use. Therefore, we employ \textbf{\emph{Header}} segment containing the metadata describing the whole partition.  We allocate memory in times of 16 bytes and address the memory space with eight-bytes-length address.  When the vector is updated, we save a new version of the vector in Data segment of the memory space and update the corresponding bytes in Index. After that, we reclaim the space that is used to save the old version. We reclaim the memory space through a set of LinkedLists in the Header segment of the off-heap memory. The offset of the first LinkedList is $RECLAIMED\_LIST$. Given a vector with the size of $s$ bytes, when its space is reclaimed, we add the offset address of the memory block to the LinkedList with the offset $RECLAIMED\_LIST + (s - 16) / 2$. When we allocate the memory space for a new vector with the size of $s$ bytes, we first check if there are reusable space in the LinkedList offsetting at $RECLAIMED\_LIST + (s - 16) / 2$.

In practice, we have a limited number of LinkedLists, thus, have the maximum size of each allocated memory block. We chain the memory blocks in order to support the vector whose size is longer than the maximum memory block size.

\subsubsection{Flash Memory}
\label{sec:storage:hierarchical_memory:flash}

Given the limited space in off-heap memory, we employ flash memory to accommodate a high volume of data in the online environments. PFO allows the user to set a threshold of the data size in off-heap space. When the data size is beyond the threshold, we make a read-only snapshot of the data in off-heap space, and save snapshot in the flash memory (referred as $S_{ij}$ in Figure \ref{fig:storage}, where $i$ is the table's partition $i$ in off-heap space, $j$ stands for the timestamp when generating this snapshot). Figure \ref{fig:storage} shows that the individual snapshot in flash memory contains \emph{Index} and \emph{Data} files, corresponding to Indexing and Data segments in Figure \ref{fig:mem_layout}.

 Given two vectors, different LSHTables that determine the near neighboring relationship among the vectors with the different set of hash functions may give different answers to whether they are in $A(q)$ when the other is the query data point. As a result, it is impossible for us to arrange two vectors in MainTable in the sequential disk space according to information in a particular LSHTable. In another word, the access pattern in MainTable is in random. We utilize flash memory to serve \textbf{query request} with the salient performance. According to the previous studies, there is no evident gap between the random and sequential reading performance in SSD-based flash memory \cite{ssdrandomread}. By utilizing the flash memory, we do not need to pay extra overhead even vectors in MainTable are put in random places. Regarding the \textbf{update request}, flash memory does not provide the random write performance that is as good as its read counterpart so that we need to avoid the random write. The snapshot we keep in flash memory layer is read-only, i.e. when off-heap partition reaches its threshold, we create the snapshot, and the future updates on the vector data is reflected in the snapshots we will create in the future instead of modifying the existing snapshot. Since we write only once for every snapshot, it is easy to perform the sequential write.

With the time passing, each partition may have multiple snapshots in flash memory. As a result, searching a particular data point requires traversing multiple snapshot files. This process brings large query overhead. To deal with multiple snapshots, we generate a summary for each snapshot with Bloom Filter \cite{bloomfilter}. The Bloom filter is a compact signature built on the hash keys in each snapshot. The key in the Bloom Filter of the MainTable is the vector ID. In the LSHTables, we use the indices of all non-empty buckets as the keys of Bloom Filters. To search $A(q)$ in flash memory, we go through the snapshot files in the reversed time order. For each snapshot, we test the Bloom filters instead of searching snapshot file directly. If any Bloom filter matches, the corresponding vectors are retrieved from flash. Bloom Filter based lookups may result in false positive; thus, a match could lead to an unnecessary flash I/O. According to studies in \cite{summarycache}, we are able to control the false positive rate as low as less than 0.001 with a very small disk space cost. Additionally, we have the periodical system maintenance routines to merge the snapshot by eliminating the duplicate vectors.

\section{Design for Parallel-Friendly Computation}
\label{sec:parallel}

Compared to the state-of-the-art parallel LSH design, PFO is more feasible to handle a large number of concurrent query and update requests and maximize system throughput for the admitted requests. PFO benefits from two design innovations: (1) an indexing structure facilitating the parallel data access with LSH-aware partitioning (Section \ref{sec:parallel:indexing}), and (2) a concurrency management module mitigating the cross-threads synchronization (Section \ref{sec:parallel:concurrency}). In this section, we will demonstrate the design details of these two innovations.

\subsection{Parallel-friendly Indexing}
\label{sec:parallel:indexing}

To enable the parallelization, we partition the data into several groups and make the access to different data points affect each other as little as possible. PFO adopts Partitioned Hash Forest (PHF) to achieve this objective and partition the memory space of a hash table on two levels, Partition and HashTree level. Our goal is to place the vectors potentially to be the nearest neighbors in the same hash tree. Therefore, the data residing in different partitions or hash trees are not possible to be involved in a query or update request, i.e. requests targeting to different hash trees are to be handled in parallel.

The partitioning algorithms in MainTable and LSHTables are different. In MainTable, given a data point vector $v$, we apply a hash function designed to minimize the hash conflict, MurmurHash3 \cite{murmurhash3}, to its vector ID. To locate the hash tree within a partition for $v$, we extract the first $m$ bits of the hash value as the tree ID. The memory space partitioning in LSHTable has more challenges than that in MainTable. We have the additional requirement to preserve the distance between the data points when partitioning the memory space, i.e. we need to ensure that only the data points in the same hash tree are possible to be the nearest neighbors. Upon achieving this, we retain the desired benefits of PHF: the irrelevant data points are indexed in different trees so that the trees can be updated in parallel and independently.

The idea of the partitioning algorithm for LSHTables in PFO is to apply the LSH functions for two times. This algorithm is inspired by the Layered LSH \cite{distributedlsh} which is used to determine the location of a server saving the data point based on its content. To determine the hash tree for a given data point vector $v$ for an LSHTable, we first apply the LSH functions associated with the LSHTable and get its LSH value $h$. We take the first $m$ bits of $h$ as the hashing tree ID in \emph{HashTree} layer (we will leave the details of Hash Tree to Section \ref{sec:online:hashtree}). However, only partitioning in HashTree level is not precise enough. For example, hash values 01111 and 00000 are significantly different with each other, but they will be indexed in the same hash tree if we set $m$ as 1. We introduce $C$ locality sensitive hashing functions to decide the location of $v$ in \emph{Partition} level. By applying these functions on $h$, only the similar hashing values are in the same region in Partition level. By introducing $C$ hashing functions in Partition level and extracting $m$ bits from data point's hashing value, we separate the memory space into $2^{C + m}$ regions each of which corresponds to a hash tree and only the similar data points are in the same tree.

The parameters $C$ and $m$ influence the throughput of PFO by allowing different level of parallelism. On the other hand, the nearest neighbors might be filtered out because they are assigned to different partitions mistakenly due to the approximate nature of LSH. In this case, the accuracy of PFO suffers from data partitioning. In Section \ref{sec:eval:tuning:partitioning}, we evaluate the impact on PFO's throughput and accuracy brought by these two parameters. We prove that under the potential problem of some data points being dumped to other partitions/trees, PFO still exhibits salient throughput and accuracy.

\subsection{Concurrency Management}
\label{sec:parallel:concurrency}

With only the parallel-friendly data structure, it is not enough to maximize the system performance. The reason is that different threads handling the query/update requests target to the same hash tree have to be synchronized. To minimize the chance of synchronization, we design the concurrency management strategy of PFO with the \emph{Actor} model \cite{actor}.

Actor is a memory-efficient entity that maintains its state. The only way to query/update the state of the actors is to send the message to them. At any moment, there is at most one thread having the access to the message queue of the actor. To achieve concurrent processing, Actor model encourages the fine-grained partitioning of state, and computation. The actor is enqueued to the task queue of a thread for processing the message when its message queue is not empty.

\begin{figure}[t]
  \begin{center}
    \includegraphics[height=2.5cm]{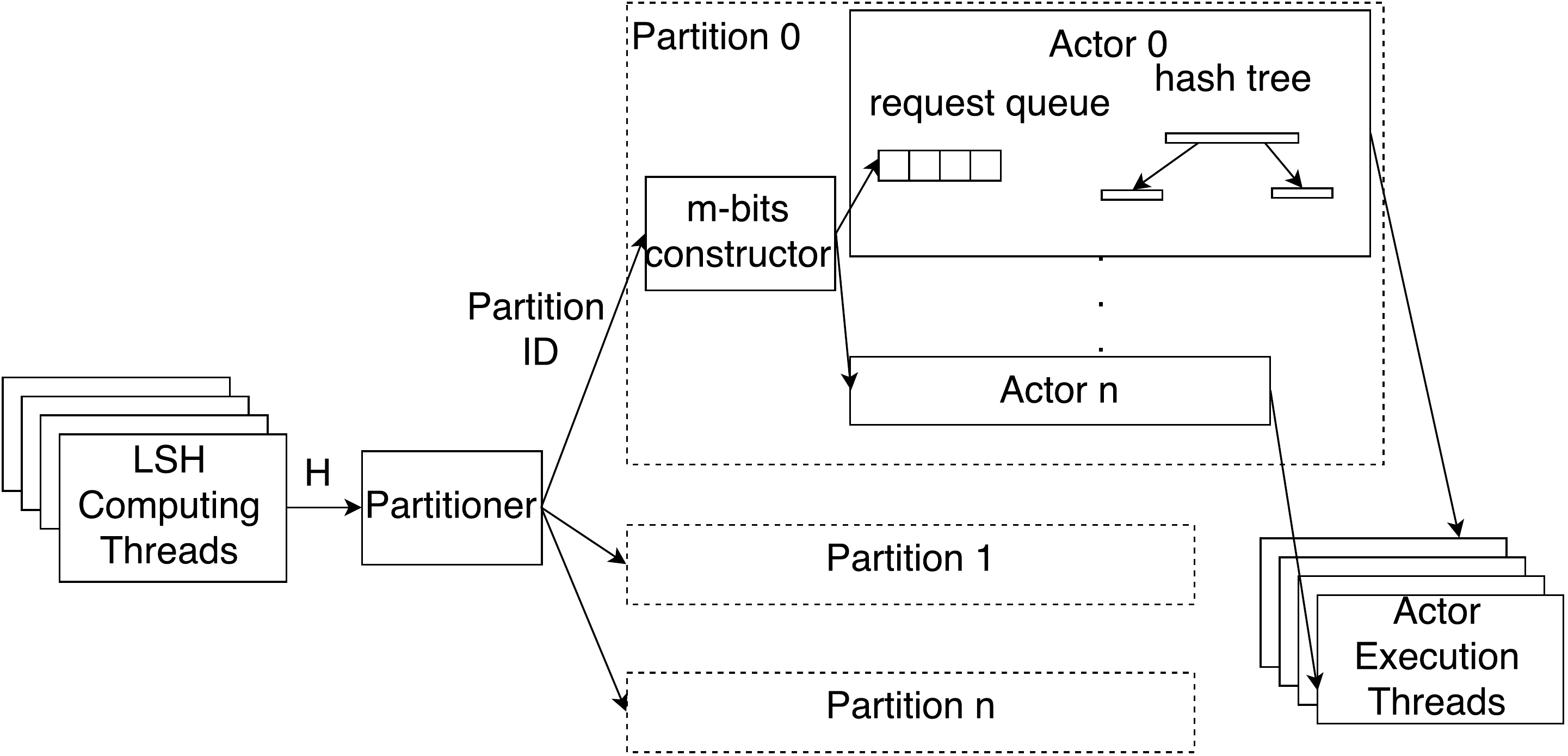}
    \caption{\small{Request Dispatching Module in PFO consists of a group of computing threads and multiple actors each of which maintains a hash tree as its state.}}
    \label{fig:actor}
  \end{center}
\end{figure}

As depicted in Figure \ref{fig:actor} the Concurrency Management module in PFO consists of a group of computing threads and multiple actors each of which maintains a hash tree as its state. The computing threads handle calculating the hash values of the query object according to LSH hash function for LSH tables or a mod-based hash function for MainTable. These threads firstly handle the requests so that the hash value calculation can be parallelized at the maximum level since there is not necessary to thread synchronization in this step. After getting the hash value $h$ of the data point $v$, we locate the hash tree for $v$ with the method we introduced Section \ref{sec:parallel:indexing}. The query/update request is then sent to the actor that maintains the corresponding hash tree as the state. After receiving the request, the actor is attached to a thread for processing the request. Actor-based concurrency module in PFO minimizes the synchronization necessary across the threads. Each thread processes the requests in its full speed instead of pausing a while just because it attempts to access some region of the hash table that is currently updated by others.

A potential problem in this design is the skewed data distribution across hash trees which causes load imbalance among actors. The similar issue appears in the general key-value store design \cite{mica, datapartition, datapartition2}. Previous studies resolve this problem well \cite{mica}. With the standard key-value store benchmark, YCSB \cite{ycsb} which exhibits a Zipf-distributed population of size 192M keys with the skewness 0.99, partitioning the memory space with the first 4 bits in hash value of the key makes the most popular partition only 53\% more frequently accessed than the average. Since we use the first $C + m$ bits to locate the hash tree, we also resolve this issue well. Additionally, multiple LSHTables in PFO will mitigate the influence of the data skewness with more hash trees. Our experimental results in Section \ref{sec:evaluation} proves that PFO keeps high throughput under this potential issue.

\section{Design for Handling Online Updates}
\label{sec:online}

The key point to accommodate the online updates is to apply the minimum change to the indexing structure when adding new data points. Operations altering the shape of the indexing structure consumes many CPU cycles and blocks the access to the affected region of the index until the reshaping is finished. For example, LSB-Tree \cite{lsbtree} adopts the B-Tree to index all data points while B-Tree itself is more read-friendly other than write-friendly due to the operations like splitting the nodes. In this section, we propose to use a hash tree structure which limits the involved range in the tree within a single slot of the non-leaf node when new data points are added and keep the query effectiveness by adjusting the resolution to identify the nearest neighbors.

\subsection{Adaptive Hash Tree}
\label{sec:online:hashtree}

\begin{figure}[t]
  \begin{center}
    \includegraphics[height=3.5cm]{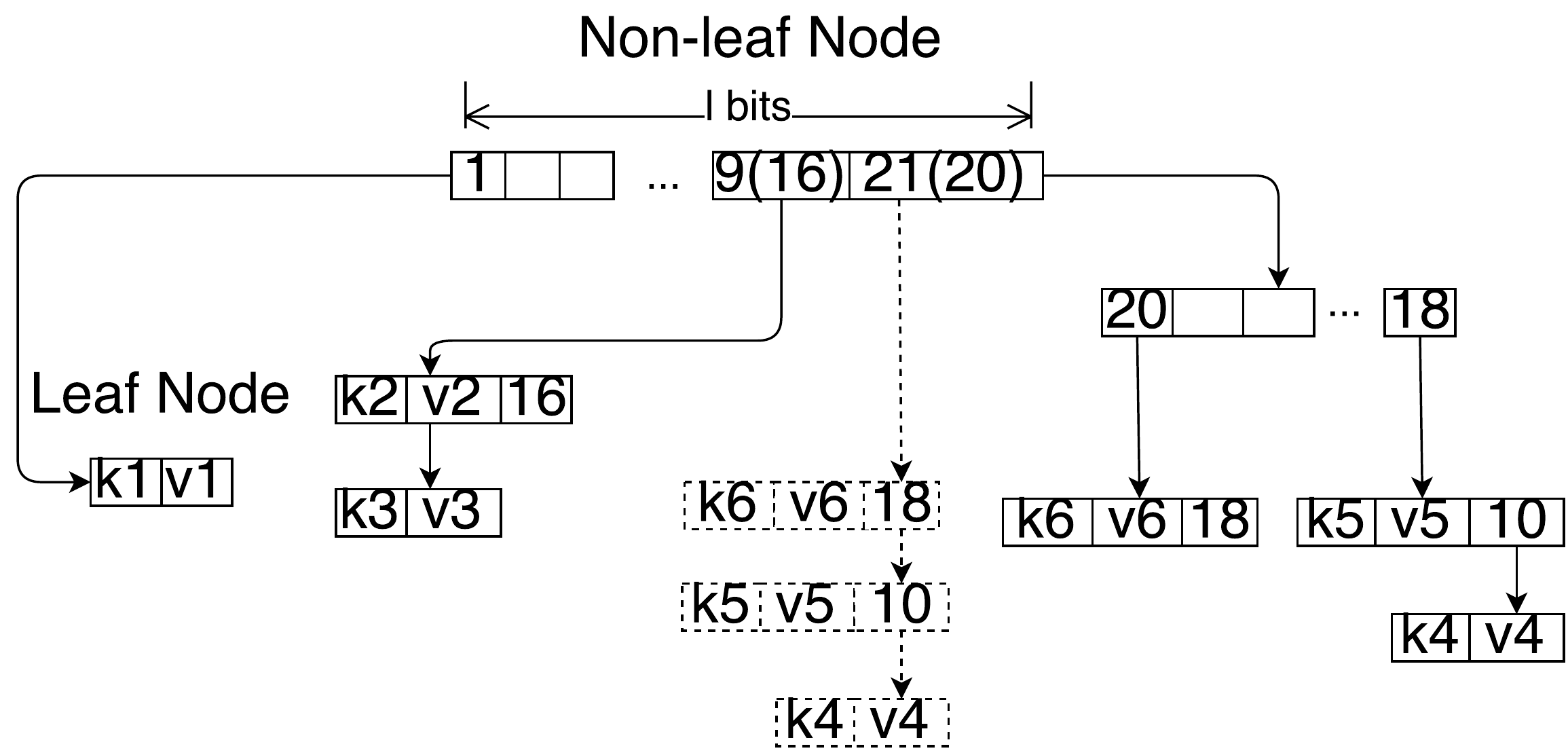}
    \caption{\small{Adaptive Hash Tree Structure: we progressively include more bits in the hash value of the vectors to locate them in the buckets. When there are more than $t$ nodes under the same buckets we automatically spread them to the next level of the hash tree.}}
    \label{fig:hash_tree}
  \end{center}
\end{figure}

Figure \ref{fig:hash_tree} illustrates the hash tree structure. The hash tree consists of two types of nodes, including leaf node and non-leaf node. The non-leaf node is logically an integer array with the length $l$. Each slot in the array maps to a bucket in the hash table. The value in the slot of the array is the offset of the first leaf node in the slot or the offset of the non-leaf node in the next level of the tree. A leaf node consists of three fields: $KEY$ field saves the vector ID, $VALUE$ field is the vector data (for MainTable) or not exists (for LSHTable), $NEXT$ indicates the offset of the next vector within the same slot of the non-leaf node.

We first discuss the update process. Given a vector and a particular LSHTable, after we locate the partition and hash tree it belongs to with the method in Section \ref{sec:parallel:indexing} (In the following algorithm, we take LSHTable as the example. MainTable is mostly the same except that the hash function associated with the table is the MurmurHash3 \cite{murmurhash3} instead of Locality Sensitive Hash functions.):

\begin{itemize}
  \item \textbf{Step 1}: We encapsulate the vector's ID as a leaf node, save the node in the off-heap space with the offset $o$. Given that the LSH value of the vector is $h$, we use the $log_2(l)$ bits following the first $m$ bits of $h$ as the vector's slot position in the root level non-leaf node;
  \item \textbf{Step 2}: If the slot has not been occupied (e.g. leaf node (K1, V1) in Figure \ref{fig:hash_tree}), we update the value in the corresponding slot of the root node with the offset of the leaf node and terminate the update process. Otherwise, we move to Step 3;
  \item \textbf{Step 3}: If the slot has been occupied, we fetch the memory block pointed by the offset currently saved in the slot of the root node. If the fetched memory block is a non-leaf node, we use the next $log_2(l)$ bits of $h$ as its slot position in this newly fetched non-leaf node. We repeat the process until we find an empty slot or we used all bits in the LSH value. If we find an empty slot, we perform the operations in Step 2. If we eventually get a slot storing the offset of a leaf node, we need to update the $NEXT$ field of the current leaf node with the offset currently saved in the slot, and then update the value in the slot of the non-leaf node as $o$. The middle part of Figure \ref{fig:hash_tree} illustrates such a scenario. (K3, V3) was originally in the root level of the hash tree, with the offset as 16. After we insert (K2, V2) which is with the offset of 9, we replace 16 in the slot of the root node with 9 and update the NEXT field of (K2, V2) as 16;
  \item \textbf{Step 4}: After writing the leaf node to off-heap memory, we examine that if there are more than $t$ nodes linked under the same non-leaf node slot. If that is the case, we spread the leaf nodes to the next level in the hash tree. We first create a new non-leaf node, update the current slot with the offset of the new non-leaf node, and move all the nodes to the new non-leaf node with the next $log_2(l)$ bits of their hash code as their slot in the new non-leaf node. The right most of Figure \ref{fig:hash_tree} describes such a case. Suppose we set $t$ as 2. After we insert (K6, V6) whose offset is 20, we found that there have been more than $t$ nodes in the same bucket. To spread these nodes to the next level, we add a new non-leaf node with the offset of 21 and replace the 20 in the slot of the root node with 21. We then extract the next $log_2(l)$ bits in the hash values of (K4, V4), (K5, V5) and (K6, V6) and use them as the slot index in the new non-leaf node.
\end{itemize}

The query process of the hash tree is similar to update, i.e. locating the vectors with every $log_2(l)$ bits in the hash value.

The fundamental idea of the above algorithm resides on two sides, adaptively control the resolution to identify nearest neighbors and minimize the involved range of indexing structure change with update requests. (1) We progressively include more bits in the hash value of the vectors to form the ID of the bucket where it locates. According to Definition 2 in Section \ref{sec:background:lsh}, the distance between two vectors is the length of the common prefix of their hash values. By setting an explicit threshold $t$, we can guarantee the effectiveness of the responses to nearest neighbors queries. Given a query data point $q$, when there are few vectors (less than $t$) have the common prefix with $q$, we want to include these data points as the candidates of the nearest neighbors to avoid return empty to the query. When there are more vectors (more than $t$) in the same bucket, we have to improve the resolution to identify the nearest neighbors. We raise the bar on the minimum length of the common prefix by moving the data points whose hash values differ in the following $log_2(l)$ bits to different buckets (Step 4). Without this step, we may have too many candidates to be checked whether they are within the distance of R from $q$. (2) When we need to change the indexing structure by redistributing the data points, we limit the involved range in a single bucket, instead of propagating the changes up until the root node, like B-Tree. This approach uses fewer CPU cycles to process online update requests by applying the minimum changes to the indexing structure.

Parameters $l$ and $t$ define the shape of the hash tree by regulating the number of data points in each bucket, and in turn influences the efficiency and accuracy of nearest neighbors search. Larger value of $t$ increases the chances to find nearest neighbors by enlarging the search range, yet introduces larger overhead for many data points to be included in similarity calculation to find the ones within the user-defined distance, R. Too small $t$ excludes too many data points probably including the ones within R from the query data point.  The change of the $l$ value exhibits the reversed impact against $t$ on the efficiency and accuracy. With the standard benchmark datasets MNIST \cite{mnist} and COLOR \cite{color}, we prove that PFO retains the salient efficiency and much higher accuracy than the competitors with parameter tuning (Section \ref{sec:eval:tuning:hashtree} and \ref{sec:eval:accuracy}).

\section{Beyond General Online Database System}
\label{sec:database}

While PFO and the general database systems share the same design goal of serving the online query/update requests, PFO is not just applying the techniques in the general database systems in the context of Locality Sensitive Hashing. In this section, we describe how Locality Sensitive Hashing makes the solutions of the online database systems not feasible in PFO, and the innovations in PFO comparing to the general online database systems.

The first difference is on the data partition. The major goal of the partitioning algorithm in the general online database systems is to distribute data evenly and maximize the parallelism of the access \cite{datapartition, datapartition2, mica}. For instance, CPHash \cite{datapartition2} creates multiple partitions for the total key space and assigns one or more partitions to a CPU core. Unfortunately, Locality Sensitive Hashing makes this approach insufficient to work well. As we indicated in Section \ref{sec:parallel:indexing}, without the additional LSH functions being applied to the hash key, irrelevant data may locate in the same hash tree, thus degrading the system performance. To address this challenge, we designed a re-locality-sensitive-hashing approach to locate only similar data in the same partition.

Another important difference is on the concurrency management. The major approaches adopted in the general online database systems fall into three categories: EREW (Exclusive Read Exclusive Write), CREW (Concurrent Read Exclusive Write) and CRCW (Concurrent Read Concurrent Write) \cite{mica}. EREW assigns a single thread for each data partition; CREW allocates a CPU core for updating request to all partitions and all the other cores for reading; CRCW simply performs the request processing with any thread in a thread pool. EREW's drawback is that it may suffer from the expensive cache line transfer due to the context switch of threads and the skewed workload. CREW and CRCW involve many read-write conflicts in the scenario of online nearest neighbors search. In contrast, PFO adopts an actor-based approach that pushes the triggered actor maintaining the hash tree to the thread task queue instead of requesting one thread per partition, hence avoiding the expensive cache line transfer. Additionally, the salient feature of exclusive single-threaded guarantee of the actor eliminates the need for synchronization and inter-core communication, making PFO scale close to linearly with CPU cores.

\section{Evaluation}
\label{sec:evaluation}

To evaluate PFO, we implemented a prototype based on a database engine, MapDB \cite{mapdb}. MapDB is a pure-Java database and we choose MapDB because of its clear interfaces and implementation, so that we can easily customize MapDB to achieve our goal. We implemented the hierarchical memory system by customizing MapDB's storage module and implemented PHF based on MapDB's hash tree implementation. Additionally, we replaced MapDB's multi-threading module with our actor-based concurrency management module. Through extensive evaluation, we show our new system improve the scalability of the original MapDB under concurrent query and update requests of nearest neighbors (subsection \ref{sec:eval:concurrency:indexing}).

In the following, we first describe our evaluation setup (subsection \ref{sec:eval:setup}) and analyze the experimental results in the next sections (subsections \ref{sec:eval:storage} - \ref{sec:eval:accuracy}).

\subsection{Evaluation Setup}
\label{sec:eval:setup}

Our testbed equips with Intel(R) Xeon(R) CPU E5-2687W v3 (25M Cache, 3.10 GHz) and 32GB RAM memory (we allocate 8 GB on-heap space to PFO JVM process). The testbed server uses SK hynix SH920 2.5 7MM 512GB SSD as the external memory. Unless mentioned otherwise, we use 10 LSHTables ($L = 10$) in PFO, each of which was partitioned into 16 partitions ($C=4$). In each partition, we use 16 hash trees ($m=4$). Within each tree, we allow at most 4 nodes in the same bucket (except the bottom level) ($t=4$), the length of non-leaf node is 128 ($l=128$) and the length of the hash value is 32 ($M=32$).

We use three datasets to evaluate PFO. The first one is Enron Email Dataset \cite{dataset} which contains  around 650,000 emails. We preprocessed the dataset with TF-IDF weighting scheme by only selecting the top weighted 0.5\% words, which limits the vector dimensionality as 9,331. We use this dataset to test the scalability and efficiency of PFO because the total number and the dimensionality of the vectors are large. We use MNIST \cite{mnist} and COLOR \cite{color} datasets to evaluate the accuracy of PFO. The MNIST dataset contains 60,000 points. Each point is a 784-dimensional vector representing the pixel value of a 28 * 28 image. The dataset also contains a test set of 10,000 points. The COLOR dataset contains 68,040 instances, each of which describes the color histogram of an image. We use these two datasets to facilitate our comparison with the competitors because they are used by most of the related work \cite{lsbtree, sklsh}. To keep consistent with the other work and make a fair comparison, we preprocessed the datasets by keeping only 50 dimensions in MNIST dataset with the largest variances, and chose 50 vectors in random from both datasets as the queries.

In the following sections, we aim to answer the following questions through the extensive evaluation:

\begin{itemize}
  \item Does the hierarchical memory design in PFO improve the system efficiency and scalability, comparing with a single-layer design? (Subsection \ref{sec:eval:storage})
  \item Does the concurrency management in PFO improve the system capability to handle online requests? How does it compare with the conventional multi-thread model and the conventional indexing structure used in the related work, e.g. B-Tree in LSB-Tree \cite{lsbtree}? (Subsection \ref{sec:eval:concurrency})
  \item How is PFO sensitive to the system parameters, i.e. how a user tunes the performance of PFO, regarding efficiency and accuracy? (Subsection \ref{sec:eval:tuning})
  \item Does the indexing structure in PFO precisely find the nearest neighbors of the query data point? (Subsection \ref{sec:eval:accuracy})
\end{itemize}

\subsection{Hierarchical Memory System in PFO}
\label{sec:eval:storage}

In this subsection, we evaluate the hierarchical memory design in PFO by measuring the query latency on each layer with various amount of data. Based on the observation of the latency change on each layer, we can get the maximum data capacity of each layer so that we know whether the hierarchical memory in PFO is superior to a single layer design.

\begin{figure}[t]
  \begin{center}
    \subfigure[\small{Minimum Read Latency of Three Types of Single-Layered Design}]{
      \includegraphics[width=0.3\textwidth]{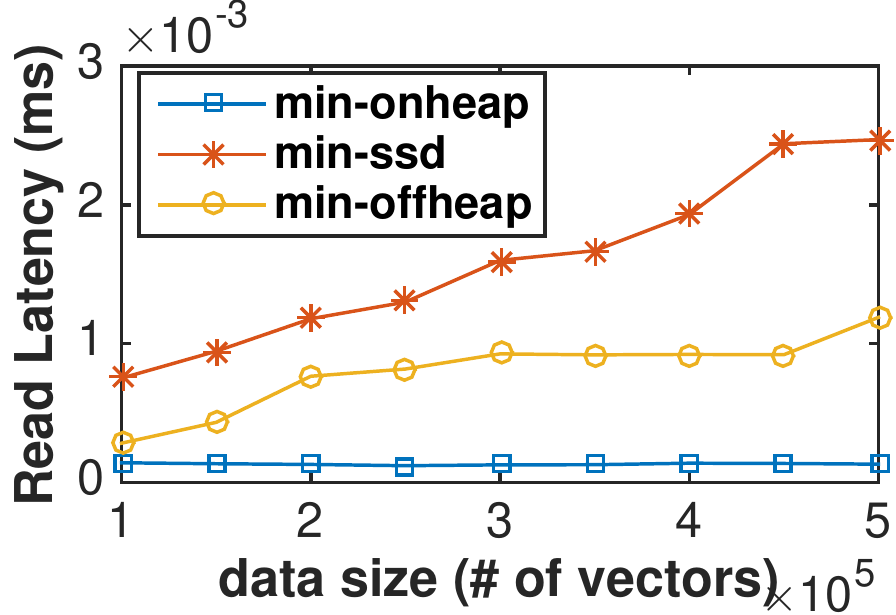}
      \label{fig:min_read}
    }
    \subfigure[\small{Average Read Latency of Three Types of Single-Layered Design}]{
      \includegraphics[width=0.3\textwidth]{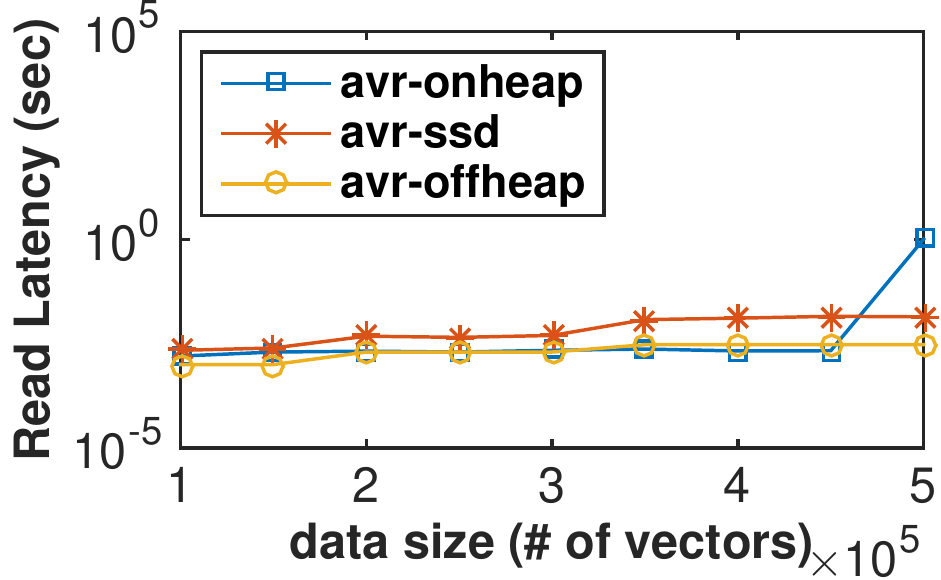}
      \label{fig:average_read}
    }
    \subfigure[\small{Max Read Latency of Three Types of Single-Layered Design}]{
      \includegraphics[width=0.3\textwidth]{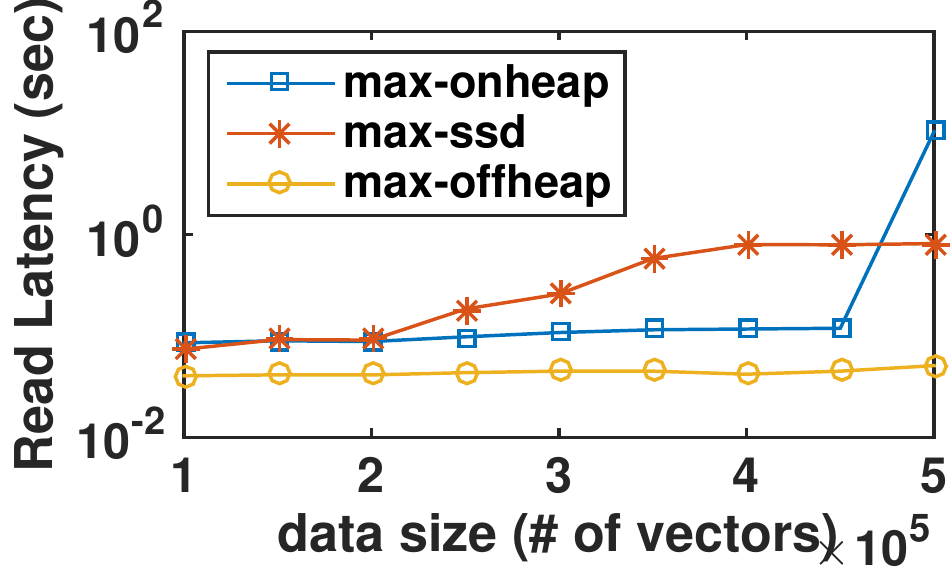}
      \label{fig:max_read}
    }
    \caption{\small{Read Latency of three types of single-layered design}}
    \label{fig:read_latency}
  \end{center}
\end{figure}

To understand the query latency in the different layers of the memory system of PFO, we run three experiments with all vector data in on-heap, off-heap and flash memory respectively. We query data in the system with ten threads, each of which generated 10,000 read requests. In these experiments, we used an Open Addressing Hash Table to index data in the on-heap case; we use PHF to index vectors in off-heap and flash memory (SSD) case. We measured the maximum, minimum and average query latency and show the results in Figure \ref{fig:read_latency}.

On average, on-heap memory space offeres the shortest read latency with no more than 450,000 vectors. The latency of off-heap and on-heap RAM are nearly the same within this data scale. Fetching the nearest neighbors for a particular query data point from SSD is around 2X slower than RAM (but still delivers sub-second latency). We observe that the trend of read latency of on-heap case is nearly flat due to the addressing method of Open Addressing Hash Table. In contrast, the latency did increase with the data size in off-heap and SSD. The reason is that it is necessary for us to deserialize bytes-represented class in off-heap and SSD memory when traversing the hash tree. As reported in \cite{sparkperf}, the CPU cost of deserialization/serialization is one of the major costs in large-scale data processing applications. More vectors indexed by PFO, the higher is the hash tree in PFO. As a result, we need to fetch more bytes from off-heap RAM/SSD and then deserialize the bytes into the directory or leaf nodes to find the nearest neighbors.

Figure \ref{fig:average_read} and \ref{fig:max_read} demonstrate the impact to the latency brought by GC. In these two figures, we observe that 1) the average latency of on-heap memory dramatically increases with 500,000 vectors; 2) the maximum latency of on-heap memory is always larger than off-heap memory. Both of the phenomenon are due to the garbage collection occurring in JVM. When GC happens (even they are just minor GC), GC threads compete with the application threads for CPU resources; the latency of the ``unlucky" requests that are processed at the same moment with GC prolong with different levels.  The worst case is that when GC module stops the application threads to struggle for more memory space, the maximum latency increased by orders of magnitude, from milliseconds to 10 seconds. In our case, it happens when we have more than 500,000 vectors. Because all the requests are blocked for waiting for GC to complete, the average latency increases accordingly, from milliseconds to second.

Through the extensive evaluation of the read latency in three layers, we have the following two observations:

\begin{itemize}
  \item \textbf{Single-Layered design is not efficient}. With a single flash memory layer, the query latency is around 2x slower than RAM; with the single layer with RAM, the capacity of the system is limited by the total size of physical RAM in the host. On-heap RAM is not a feasible solution for storing a large volume of objects as the GC would degrade performance dramatically.
  \item \textbf{PFO overcomes the drawbacks with hierarchical memory storage}. By consuming the I/O in off-heap RAM, we can process the request in memory. We scale the system capacity with SSD-based flash memory. When querying data in SSD-based flash memory, we still achieve the sub-second query latency.
\end{itemize}

\subsection{Performance With Concurrent Query/Update Requests}
\label{sec:eval:concurrency}

In this subsection, we test the scalability of the concurrency management strategy in PFO in multi-core environments. We conduct two groups of experiments to support our conclusion. First, we compare the multi-core scalability of PHF structure with a B-Tree based indexing structure as well as the ordinary MapDB hash table structure; Second, we develope a variation of PFO with the conventional concurrency management method where the request is handled by a random thread in a thread pool and compare its read/write requests throughput with PFO. We put all data in off-heap RAM in these experiments.

\subsubsection{Scalability of Indexing Structure}
\label{sec:eval:concurrency:indexing}

We first evaluate the ``parallel-friendness" of PHF indexing structure. We evaluate it by testing the read and write throughput. When we test the read performance, we choose 200,000 vectors from the dataset in random to preload to PFO and feed PFO with continuous read requests. When testing the write performance, we write 500,000 vectors to PFO. The experiments are repeated for five times and report the average value. The Core Number in X-axis represents the number of threads we used in the experiments. We choose B-Tree as the baseline for the experiments. B-Tree is one of the most widely used indexing structures in the general database engine, and it was also adopted in other related work \cite{lsbtree}. We set the node size in B-Tree as 32. Because our implementation is based on MapDB \cite{mapdb}, we also compare the throughput with the original MapDB hash table implementation showing how much we improve over MapDB.

\begin{figure}[t]
  \begin{center}
    \includegraphics[height=2.5cm]{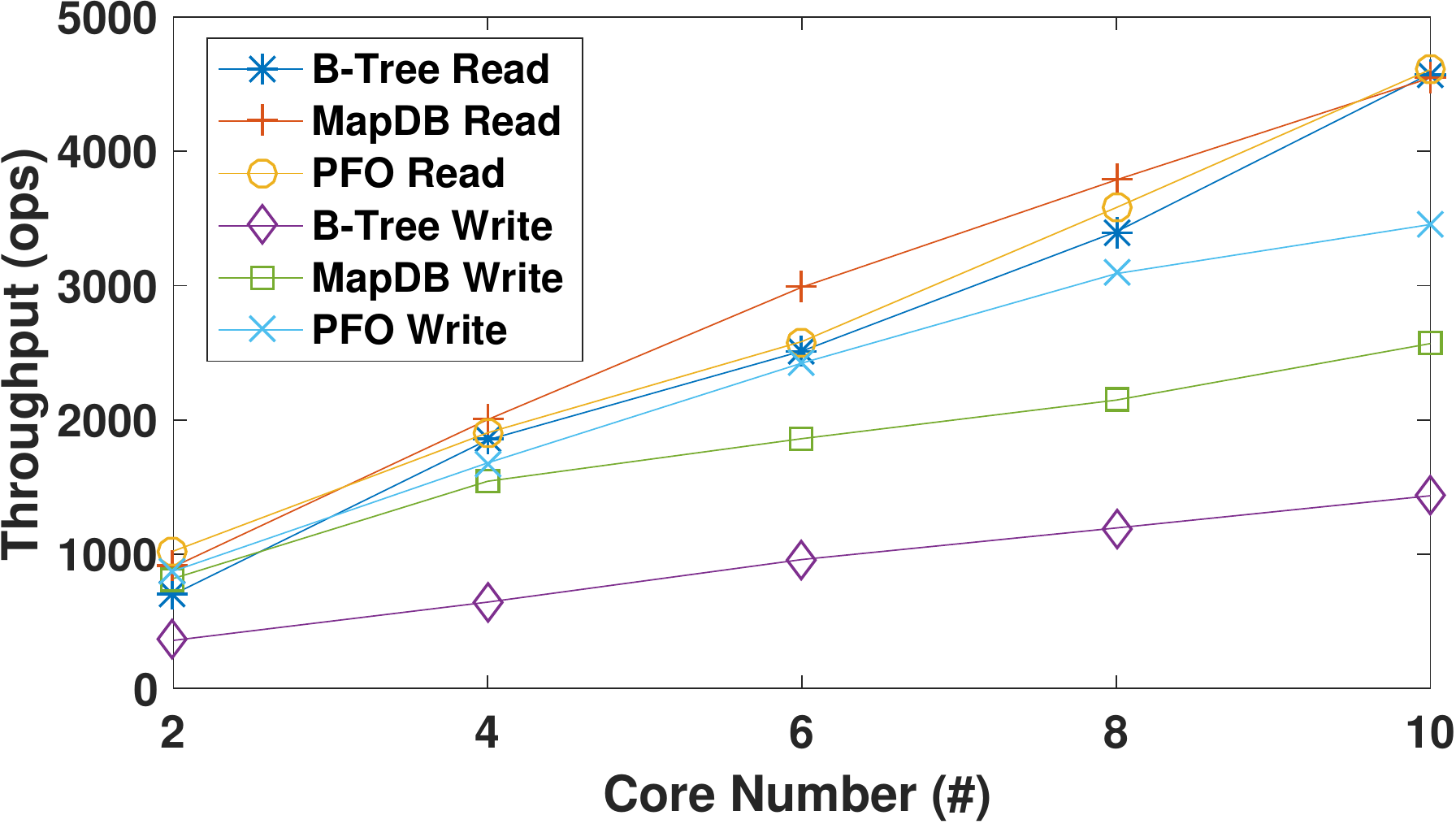}
    \caption{\small{Scalability of PFO and Other Indexing Structure.}}
    \label{fig:structure_comparison}
  \end{center}
\end{figure}

We show the experimental results in Figure \ref{fig:structure_comparison}.  The changing trends of the read throughput are identical with three structures. Because there is no synchronization across threads, and the major cost is on system cost, like context switch. On the other hand, the write throughput of MapDB and B-Tree is much lower than PFO. B-Tree suffers from the tree transformation as well as the intensive competition of locks across threads when handling write requests. The original MapDB hash tree implementation does not scale with the increasing number of cores, either. Although MapDB improves the write performance by mitigating the hash tree transformation comparing with B-Tree, the synchronization across the threads still takes the throughput down.

\subsubsection{Scalability of Concurrency Management}
\label{sec:eval:concurrency:dispatching}

To understand how much update dispatching contributes to the scalability of PFO, we develop the variation of PFO with the conventional concurrency management method where we allow concurrent threads access any data in the index. We evaluated the read/write throughput with the same workload as in Section \ref{sec:eval:concurrency:indexing}.

\begin{figure}[t]
  \begin{center}
    \includegraphics[height=2.5cm]{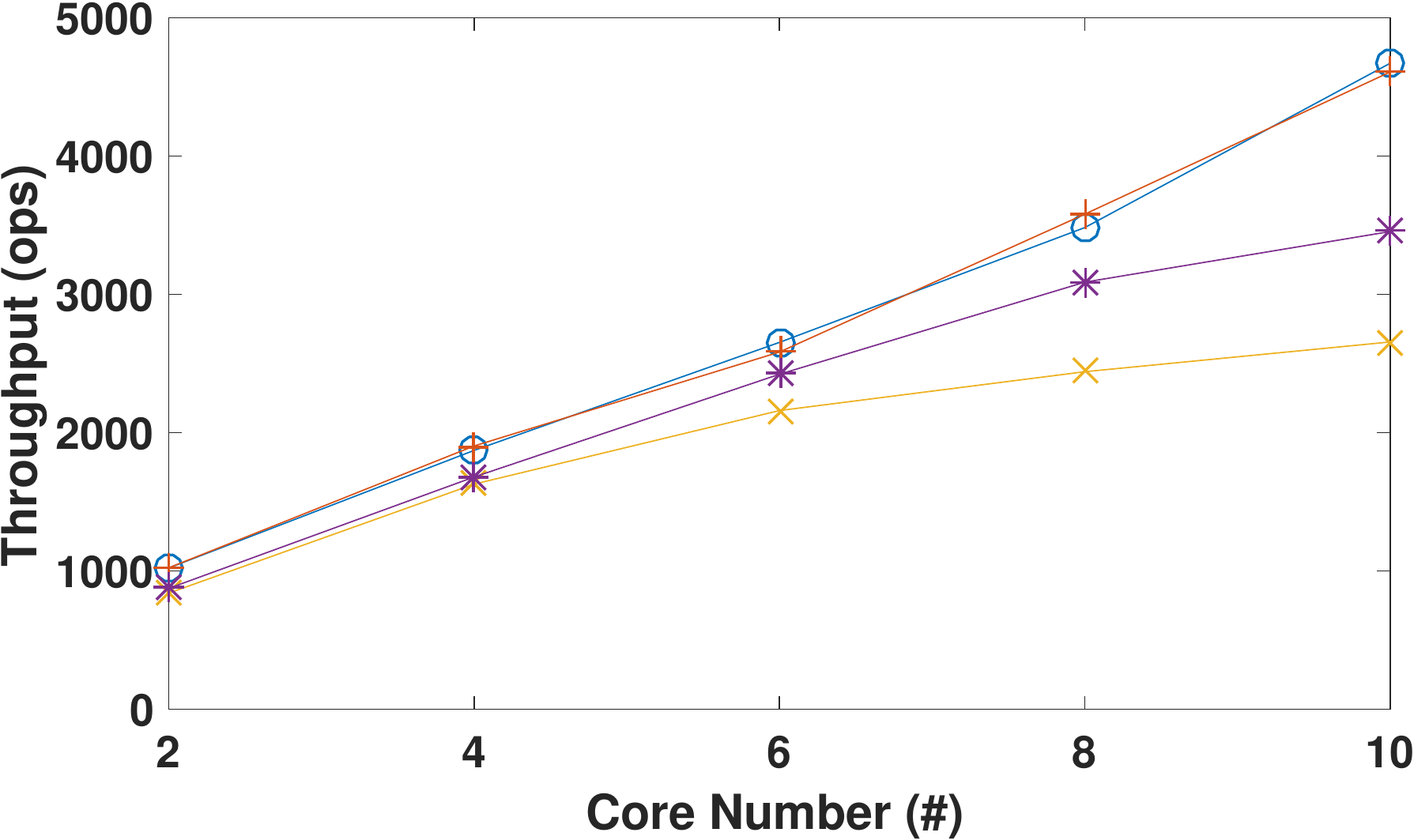}
    \caption{\small{Scalability of PFO against Core Number}}
    \label{fig:perf_concurrency}
  \end{center}
\end{figure}

With the increasing number of cores we allocate to PFO in the experiments, we find that the read throughput increases in the same pattern under both of these two concurrency management strategies. Regarding the write performance curves in Figure \ref{fig:perf_concurrency}, we find that the performance improvement starts slows down when we have 6 cores under the conventional concurrency management strategy. By comparing the write performance of PFO and its variance based on conventional concurrency management strategy, we observe that the concurrency management module in PFO improves the throughput with 30\%.

\subsection{Parameter Sensitivity}
\label{sec:eval:tuning}

In this subsection, we discuss the sensitivity of PFO against different parameters. The first type of parameters is the parameters regulating the number of data partitions, i.e. $C$ and $m$. The second type are the parameters deciding the shape of each hash tree, i.e. $t$ and $l$ which regulate how many elements are allowed to be in the same bucket (above the last level of the tree) and the size of the directory node in the hash tree. In this section, we focus on the performance of a single LSHTable where all of these parameters apply impact.

\subsubsection{Data Partitioning Parameters}
\label{sec:eval:tuning:partitioning}

\begin{figure}[t]
  \begin{center}
    \subfigure[Throughput with different $C$ and $m$]{
      \includegraphics[width=0.2\textwidth]{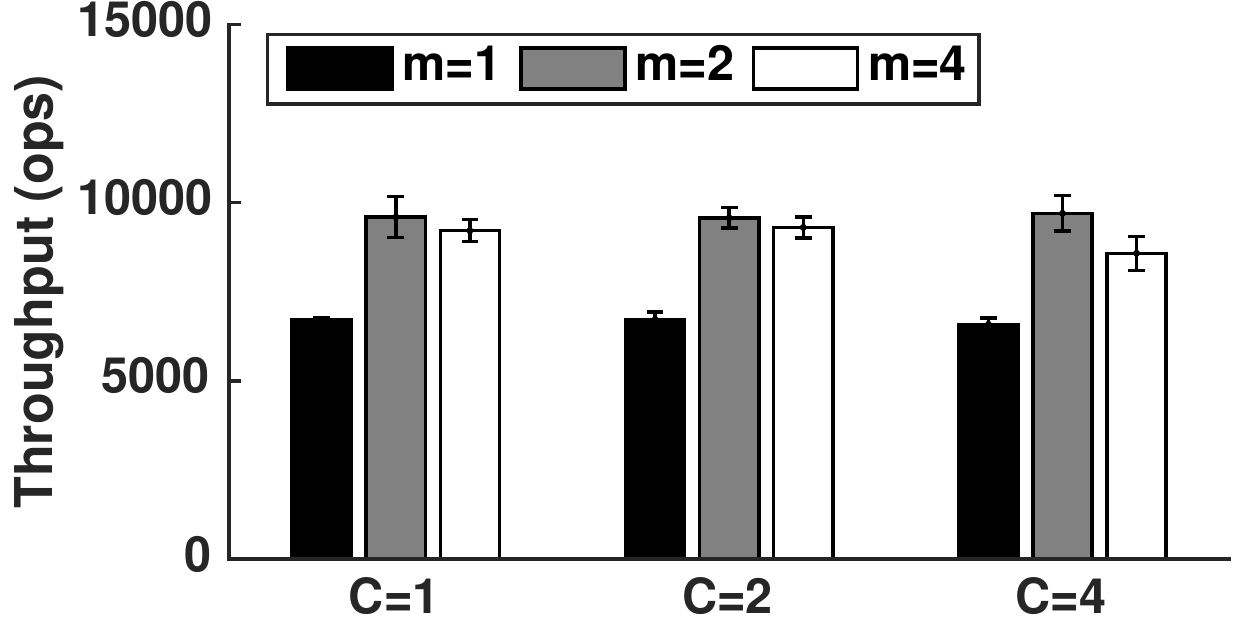}
      \label{fig:throught_m_and_c}
    }
    \subfigure[Accuracy with different $C$ and $m$]{
      \includegraphics[width=0.2\textwidth]{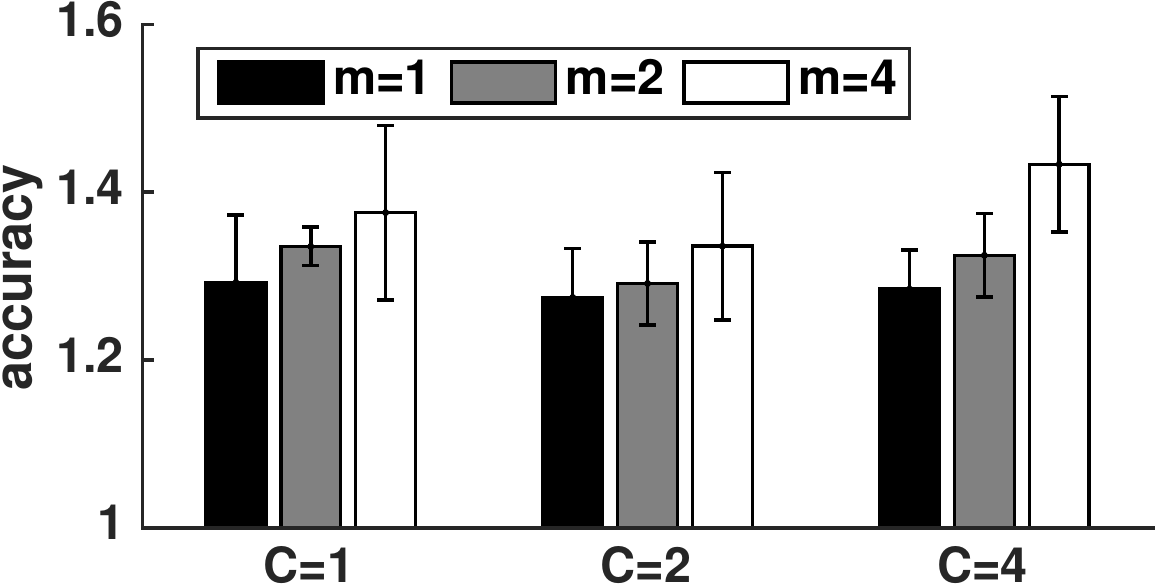}
      \label{fig:accuracy_m_and_c}
    }
    \caption{\small{System Performance Against different $C$ and $m$}}
    \label{fig:m_and_c}
  \end{center}
\end{figure}

Recall $C$ represents the number of hash functions to determine data point's partition, and $m$ is the number of bits in the data point's LSH value to decide which hash tree it belongs to. To understand how parameters $C$ and $m$ bring impacts to PFO's performance, we measure PFO's throughput and accuracy against different combinations in Figure \ref{fig:m_and_c}.  Figure \ref{fig:throught_m_and_c} shows the system throughput under a synthetic workload. We compose the workload by using ten threads loading 500,000 vectors to PFO and after saving each vector to PFO, we also search the $A(q)$ for the newly saved vector. In Figure \ref{fig:throught_m_and_c}, we observe that when we increase $m$ from 1 to 2, the system throughput significantly improved. However, when we increase $m$ to 4, the system throughput did not improve accordingly but slightly decreased. The reason is that we have only ten threads in our server while we have at least 16 hash trees in each partition. Additionally, we need to pay the cost on context switch when we have more hash trees. The other interesting observation is that increasing $C$ did not improve throughput. We investigate the reason by looking at the number of vectors in each partition when we increase $C$. It turned out that there was some data skew in the level of partition. In this experiment, we have a single LSHTable, which amplifies the negative impact of data skew. \textbf{\emph{When we have more data tables, all threads would be fully utilized since they are partitioned in different hash functions.}} The result showing in Figure \ref{fig:perf_concurrency} where PFO scales with the number of cores proves our conclusion.

Figure \ref{fig:accuracy_m_and_c} shows the accuracy of PFO brought by partitioning the hash table into multiple partitions/hash trees. Because it is hard to find a general threshold value serving well for all queries \cite{lsbtree}, we used the accuracy metric, error ratio, used in \cite{lsbtree, sklsh} to measure the accuracy of PFO. The metric is defined as:

\begin{equation}
r = \frac{1}{k}\sum_{i=1}^{k} \frac{\lVert \mathbf{o_i, q} \rVert}{\lVert \mathbf{o_i^*, q} \rVert}
\end{equation}

where $o_i$ ($i = 1...k$) are the $k$ objects found in the same bucket with $q$, $o_i^*$ ($i = 1...k$) are the ground truth of $q$'s k nearest neighbors. Both $o_i$ and $o_i^*$ are ranked by the increasing order of their distance to q. By finding k nearest neighbors for each query, we are actually setting the threshold R as the distance from the k-th neighbor to the query object. We set $k$ as 10. When less than $k$ data points are returned due to the data partitioning, we took similarity as $0$ to apply the penalty to PFO. The ideal value of $r$ is 1.

We use the standard MNIST dataset in the experiment. From Figure \ref{fig:accuracy_m_and_c}, we observe that introducing more partitions/hash trees does degrade the accuracy. In practice, we need to make the tradeoff between the accuracy and system throughput when using PFO. On the other side, the filtered out data points can be compensated by using more hash tables so that we still gain the satisfactory accuracy. Even with a single hash table, the accuracy of PFO is significantly better than the competitor. We will show our comparison in Section \ref{sec:eval:accuracy}.

\subsubsection{Hash Tree Shaping Parameters}
\label{sec:eval:tuning:hashtree}

We also test the how the shape of a hash tree influences the accuracy and efficiency of PFO. We investigate by changing the value of $l$ and $t$, and fix $m$ and $C$ as $2$ and $1$ respectively.

\begin{figure}[t]
  \begin{center}
    \subfigure[Efficiency with different $l$ and $t$, the lower the better]{
      \includegraphics[width=0.2\textwidth]{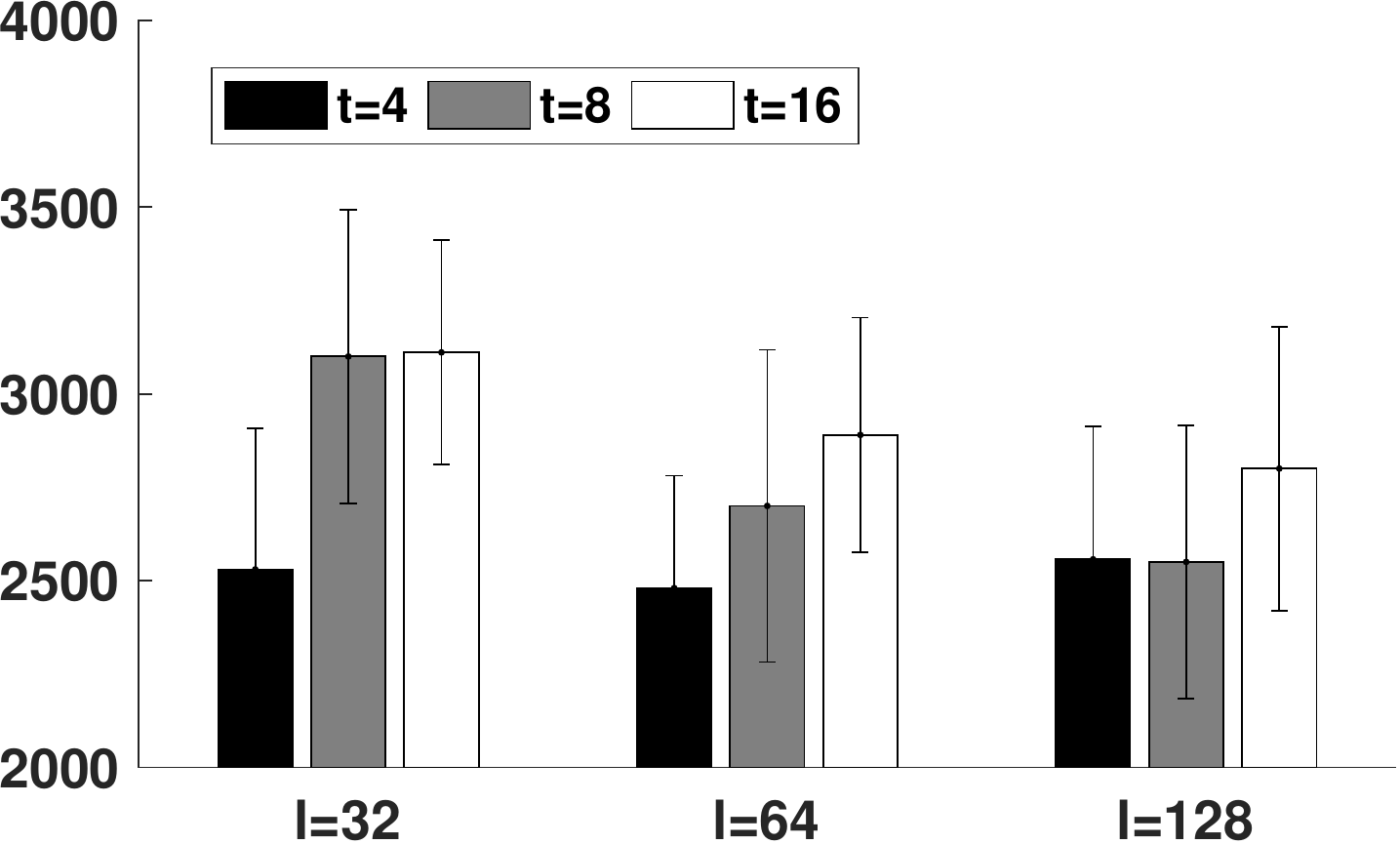}
      \label{fig:read_l_and_t}
    }
    \subfigure[Accuracy with different $l$ and $t$]{
      \includegraphics[width=0.2\textwidth]{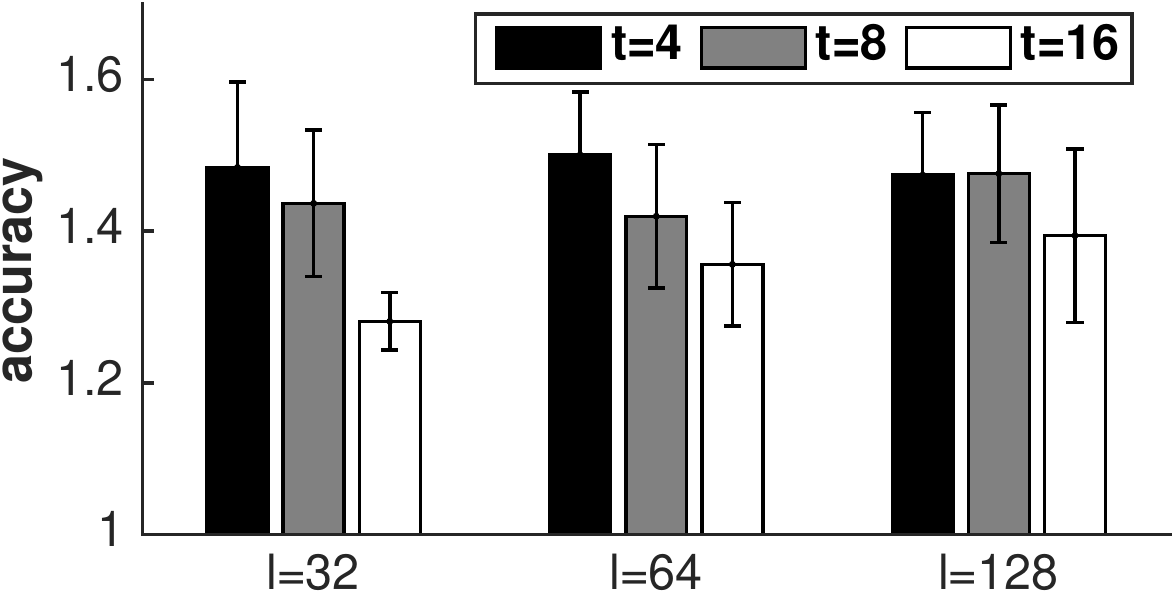}
      \label{fig:accuracy_l_and_t}
    }
    \caption{\small{System Performance Against different $l$ and $t$}}
    \label{fig:l_and_t}
  \end{center}
\end{figure}

$l$ and $t$ influence both the efficiency and accuracy of PFO. We used the accuracy metric defined in Equation (1). To define the efficiency, we measured how many data points are included in the similar search phase but has to be excluded since they are far more away with the query data point than the k-th nearest neighbors. We define the metric as following:

\begin{equation}
e = \frac{|A(q)|}{k} \quad if |A(q)| > k
\end{equation}

where $|A(q)|$ is the number of the data points which are in the same bucket with $q$. We use MNIST dataset and send 50 requests in each experiment. We demonstrate the sum of $e$ for these 50 requests in Figure \ref{fig:read_l_and_t}. Given the fixed $l$, increasing $t$ to allow more data points in the directory node introduces more overheads when searching the nearest neighbors. However, it increases the chances to find kNN by enlarging the search range as shown in Figure \ref{fig:accuracy_l_and_t}.

\subsection{Accuracy}
\label{sec:eval:accuracy}

\begin{figure}[t]
  \begin{center}
    \includegraphics[height=2.5cm]{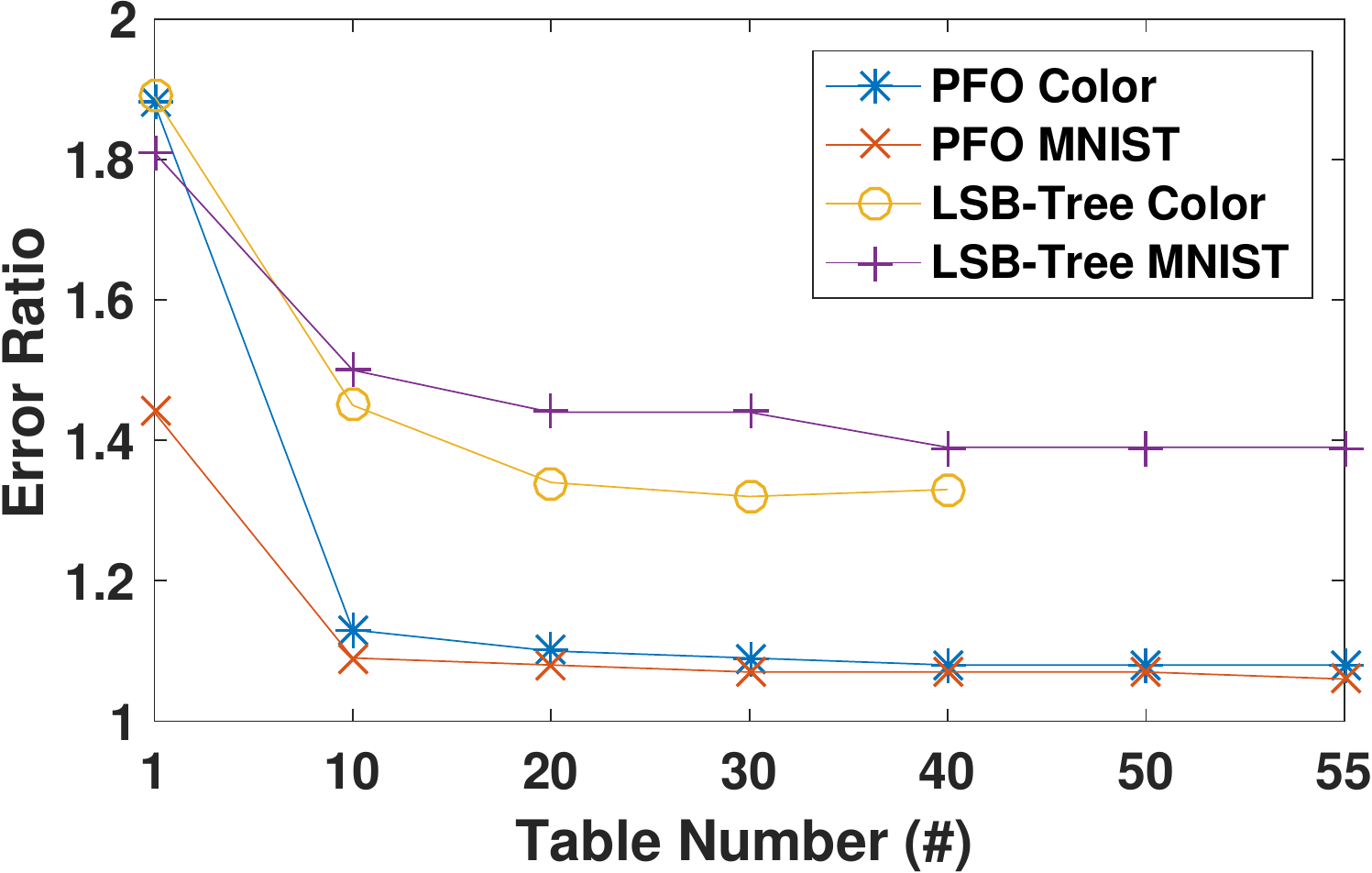}
    \caption{\small{Accuracy of PFO and LSB-Tree (k = 10)}}
    \label{fig:accuracy}
  \end{center}
\end{figure}

In order to evaluate the accuracy of PFO, we conduct to evaluate PFO is to test its accuracy. We compare the accuracy of PFO and LSB-Tree \cite{lsbtree} with the metric in Equation 1. In Figure \ref{fig:accuracy}, we use the results from the original paper describing LSB-Tree directly. We observe that the accuracy of PFO outperforms LSB-Tree with a considerable level. We attribute the improvement in accuracy to that we precisely use the hash code to index the elements while LSB-Tree converts the hash key to z-order value to be indexed by a B-Tree with the cost of accuracy loss.

\section{Related Work}
\label{sec:related}

\emph{Approximate Nearest Neighbor}: There are a large number of methods proposed for LSH-based ANN search. The primary LSH method was first suggested by Indyk and Motwani \cite{lsh}, but it is too space-consuming. The LSH functions for $l_p$ norms are discovered by Datar et al. \cite{basiclsh}. Several heuristic variations of LSH have been proposed to improve the performance of LSH-based NN search system. We summarize the existing work on the new variations of LSH in Table \ref{tab:lsh_related}. Lv et al. \cite{multiprobelsh} proposed to reduce the requested number of hash tables of LSH by probing more data within a table to answer a query. Their ideas of taking the data points with the ``similar", instead of strictly identical, hash values as NN candidates were widely adopted in the following research work.  LSB-Tree \cite{lsbtree}, SK-LSH \cite{sklsh} and C2LSH \cite{lshcollisioncounting} followed this idea with various approaches to improve the efficiency and accuracy of LSH-based systems. All of these approaches prioritize the efficiency of a single query while did not consider how to optimize the LSH indexing structure to facilitate the query/update in a concurrent environment.

\emph{Parallel AND Distributed LSH}: PLSH \cite{plsh} and Distributed LSH system proposed in \cite{distributedlsh} scale the LSH-based NN search system in a distributed fashion. Though the distributed computing schema scales the overall capacity of the system, it is only an efficient solution to improve the system scalability when we have exploited the full potentials of a single host. ``Fully utilize a single before you go to multiple ones" is also a new common agreement in the community \cite{singlehost1, singlehost2, singlehost3}. However, the proposed distributed LSH systems did not explore the full potentials of the computing power of a single host. For instance, PLSH only considers using RAM to store LSH tables, and it must broadcast a single request to all machines to fetch the NN results due to the lack of an indexing structure. Hypercurves \cite{cpugpulsh} and GPU-LSH \cite{gpuknn} utilized GPU to parallelize the index construction but it does not support to update the indexing structure in real-time.

\emph{Near Neighbours Search via LSH}: LSH is not necessary to be used to find the closest point to the query object in the feature space, thus being helpful to reduce search range for applications involving similarity search. For example, Google utilizes LSH \cite{googlenews} to cluster the users before calculating the similarities among the users for news recommendation service. LSH in Google News are calculated with MapReduce \cite{mapreduce}, i.e. only being calculated in batching but does not fit in the scenario that read/write requests arrive continuously. PLSH proposed in \cite{plsh} is to streaming the similarity search among the Tweets and LSH is also as a pre-stage minimizing the search scope before calculating the similarities. As stated above, it does not fully utilize the computing power of a single host, e.g. the capacity of the system is limited by the physical RAM size of a host, and it does not maximize the individual host throughput with the concurrent query and update requests.

\section{Conclusion}
\label{sec:conclusion}
Locality Sensitive Hashing (LSH) accelerates nearest neighbors search in high dimensional space. However, the practical system implementation is missing to support the large volume of online user requests. This paper develops a new system called PFO that provides higher throughput, better scalability and the capability to accommodate inbound data in real-time to make them accessible to the queries. It dramatically improves the space and temporal efficiency of the LSH implementations, without compromising the query response quality. It is faster than the state of the art with single-memory-layer by 2X and offers 5X higher throughput than the conventional indexing structure. Compared to the existing LSH indexing structure, our technique requires only a small portion of the query overhead to produce results in considerably better quality. Furthermore, PFO follows the ordinary key-value store schema and can be easily incorporated in the key-value storage systems.

Though we did not introduce much about fault-tolerance feature in PFO, the fault-tolerance techniques in the general database system,  e.g. with a Write-Ahead-Log, can be introduced here easily.

{\footnotesize
\bibliographystyle{abbrv}
\bibliography{draft}}
\end{document}